\documentclass[german,english]{bvm2020}

\def\articlenumber{0000}

\date{}
\title{Field of View Extension in Computed Tomography Using Deep Learning Prior}

\titlerunning{FOV Extension in CT Using Deep Learning Prior}

\author{Yixing~Huang$^1$, Lei~Gao$^1$, Alexander~Preuhs$^1$, Andreas~Maier$^{1,2}$}

\authorrunning{Huang et al.}
\institute{%
$^1$Pattern Recognition Lab, Friedrich-Alexander-University Erlangen-Nuremberg\\
$^2$Erlangen Graduate School in Advanced Optical Technologies (SAOT)
}

\email{yixing.yh.huang@fau.de}

\usepackage{subfigure}
\begin{document}

%==============================================================================
% wählen Sie mit dem Befehl \selectlanguage die Sprache aus, in der Ihr 
% Proceeding verfasst ist
%
%\selectlanguage{german}
\selectlanguage{english}

\maketitle

\begin{abstract}
In computed tomography (CT), data truncation is a common problem. Images reconstructed by the standard filtered back-projection algorithm from truncated data suffer from cupping artifacts inside the field-of-view (FOV), while anatomical structures are severely distorted or missing outside the FOV. Deep learning, particularly the U-Net, has been applied to extend the FOV as a post-processing method. Since image-to-image prediction neglects the data fidelity to measured projection data, incorrect structures, even inside the FOV, might be reconstructed by such an approach. Therefore, generating reconstructed images directly from a post-processing neural network is inadequate. In this work, we propose a data consistent reconstruction method, which utilizes deep learning reconstruction as prior for extrapolating truncated projections and a conventional iterative reconstruction to constrain the reconstruction consistent to measured raw data. Its efficacy is demonstrated in our study, achieving small average root-mean-square error of 24\,HU inside the FOV and a high structure similarity index of 0.993 for the whole body area on a test patient's CT data.

\end{abstract}

\section{Introduction}
In computed tomography (CT), image reconstruction from truncated data occurs in various situations. In region-of-interest (ROI) imaging, also known as interior tomography, collimators are inserted between the X-ray source and the detector of a CT scanner for low dose considerations. In addition, due to the limited detector size, large patients cannot be positioned entirely inside the field-of-view (FOV) of a CT scanner. In both scenarios, acquired projections are laterally truncated. Images reconstructed by the standard filtered back-projection (FBP) algorithm from such truncated data suffer from cupping artifacts inside the FOV, while anatomical structures are severely distorted or missing outside the FOV.

So far, many approaches have been investigated for truncation correction. Among them, a major category of methods are based on heuristic extrapolation, including symmetric mirroring, cosine or Gaussian functions, and water cylinder extrapolation (WCE) \cite{hsieh2004novel}. Such extrapolation methods seek for a smooth transition between measured and truncated areas to alleviate cupping artifacts. Another category of methods seek for an alternative to the standard FBP method, where the high-pass ramp filter is the main cause of cupping artifacts. Decomposing the ramp filter into a local
Laplace filter and a nonlocal low-pass filter \cite{xia2013towards} is one of such methods. Another strategy is the differentiate back-projection (DBP) \cite{Noo2004Two} approach, one milestone for interior tomography. With DBP, theoretically exact solutions have been developed based on \textit{a priori} knowledge \cite{kudo2008tiny}. With the development of compressed sensing technologies, iterative reconstruction with total variation (TV) regularization \cite{yu2009compressed} is a promising approach for interior tomography, despite its high computation.

Recently, deep learning has achieved impressive results in various CT reconstruction fields \cite{wang2018image}, including low-dose denoising \cite{wolterink2017generative,kang2018deep,yang2018low}, sparse-view reconstruction \cite{chen2018learn}, limited angle tomography \cite{wurfl2018deep}, and metal artifact reduction \cite{zhang2018convolutional}. In the field of interior tomography, Han and Ye applied the U-Net to remove null space artifacts \cite{schwab2019deep} from FBP reconstruction. Observing its instability, they propose to use DBP reconstruction instead of the FBP reconstruction as the input of the U-Net for various types of ROI reconstruction tasks \cite{han2018one}. Except for learning-based post-processing methods, interior tomography images can be directly learned from truncated data by the iCT-Net \cite{li2019learning} based on known operators \cite{maier2019learning}. For FOV extension, Fourni{\'e} et al. \cite{fournie2019ct} have demonstrated the efficacy of the U-Net in this application. However, no thorough evaluation is provided in their preliminary results.

Although deep learning surpasses conventional methods in many CT reconstruction fields, its robustness remains a concern for clinical applications \cite{huang2018some}. Since post-processing neural networks have no direct connections to measured projection data, incorrect structures, even inside the FOV, might be reconstructed. Therefore, generating reconstructed images directly from a post-processing neural network is inadequate. In this work, we propose a data consistent reconstruction (DCR) method \cite{huang2019data} to improve the image quality of deep learning reconstruction for FOV extension. It utilizes deep learning reconstruction as prior for data extrapolation and a conventional iterative reconstruction method to constrain the reconstruction consistent to measured projection data.

\section{Materials and Methods}
Our proposed DCR method consists of three main steps: deep learning artifact reduction, data extrapolation using deep learning prior, and iterative reconstruction with TV regularization.
\subsection{Deep Learning Artifact Reduction}
\begin{figure}[tbh]
\centering
\includegraphics[width = 1\textwidth]{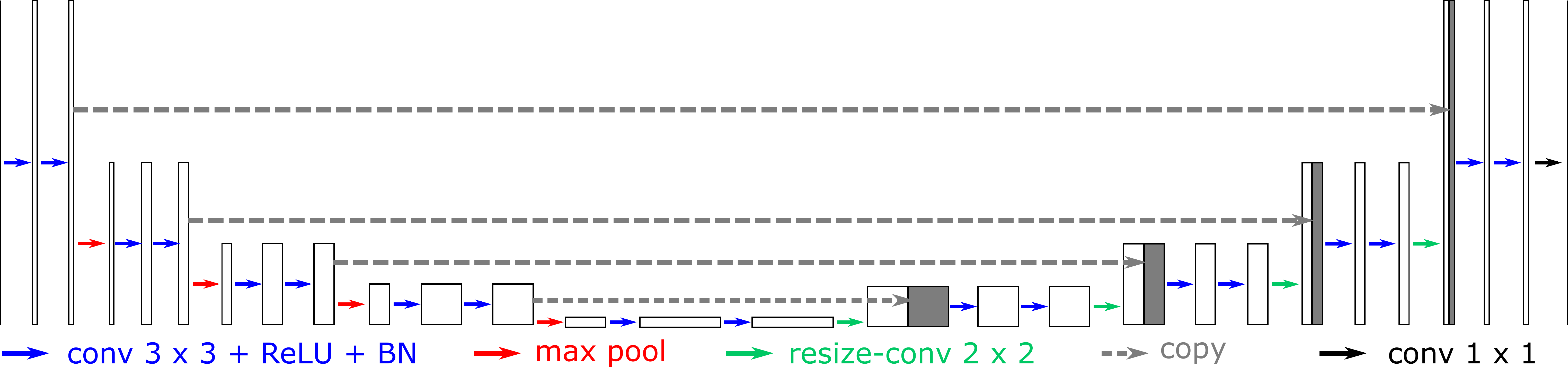}
\caption{The U-Net architecture for truncation artifact reduction.}
\label{Fig:UNetArch}
\end{figure}
As displayed in Fig.~\ref{Fig:UNetArch}, the state-of-the-art U-Net is used for truncation artifact reduction. Images reconstructed by FBP directly from truncated projections suffer from severe cupping artifacts, especially at the FOV boundary. It is difficult for the U-Net to learn the corresponding artifacts accurately, according to our experiments. Instead, FBP reconstruction from extrapolated projections contains much fewer cupping artifacts. Therefore, in this work, an image reconstructed from WCE \cite{hsieh2004novel} processed projections, denoted by $\boldsymbol{f}_{\text{WCE}}$, is chosen as the input of the U-Net. The output of the U-Net is its corresponding artifact image, denoted by $\boldsymbol{f}_{\text{artifact}}$. Then an estimation of the artifact-free image, denoted by $\boldsymbol{f}_{\text{U-Net}}$, is obtained by $\boldsymbol{f}_{\text{U-Net}} = \boldsymbol{f}_{\text{WCE}} - \boldsymbol{f}_{\text{artifact}}$. 

\subsection{Data Extrapolation Using Deep Learning Prior}
For data consistent reconstruction, we propose to preserve measured projections entirely and use the deep learning reconstruction as prior for extrapolating missing (truncated) data. We denote measured projections by $\boldsymbol{p}_{\text{m}}$ and their corresponding system matrix by $\boldsymbol{A}_{\text{m}}$. We further denote truncated projections by $\boldsymbol{p}_{\text{t}}$ and their corresponding system matrix by $\boldsymbol{A}_{\text{t}}$. The deep learning reconstruction $\boldsymbol{f}_{\text{U-Net}}$ provides prior information for the truncated projections $\boldsymbol{p}_{\text{t}}$. Therefore, an estimation of $\boldsymbol{p}_{\text{t}}$, denoted by $\hat{\boldsymbol{p}}_{\text{t}}$, is achieved by forward projection of $\boldsymbol{f}_{\text{U-Net}}$,
\begin{equation}
\hat{\boldsymbol{p}}_{\text{t}} = \boldsymbol{A}_{\text{t}} \boldsymbol{f}_{\text{U-Net}}.
\end{equation} 
Combining $\hat{\boldsymbol{p}}_{\text{t}}$ with $\boldsymbol{p}_{\text{m}}$, a complete projection set is obtained for extended FOV reconstruction.

\subsection{Iterative Reconstruction with TV Regularization}
Due to intensity discontinuity between $\hat{\boldsymbol{p}}_{\text{t}}$ and $\boldsymbol{p}_{\text{m}}$ at the transition area, artifacts occur at the boundary of the original FOV in the image reconstructed directly by FBP. Therefore, iterative reconstruction with reweighted total variation (wTV) regularization is utilized,
\begin{equation}
 \min||\boldsymbol{f}||_{\text{wTV}}, \text{ subject to } \left\lbrace
 \begin{array}{l}
 ||\boldsymbol{A}_{\text{m}}\boldsymbol{f} - \boldsymbol{p}_{\text{m}}|| < e_1,\\
 ||\boldsymbol{A}_{\text{t}}\boldsymbol{f} - \hat{\boldsymbol{p}}_{\text{t}}|| < e_2.
 \end{array} 
 \right.
 \label{eqn:ObjectiveDataConsistentDeepLearning}
 \end{equation}

Here $e_1$ is a noise tolerance parameter for the data fidelity term of the measured projections and the other tolerance parameter $e_2$ accounts for the inaccuracy of the deep learning prior $\boldsymbol{f}_{\text{U-Net}}$. $||\boldsymbol{f}||_{\text{wTV}}$ is an iterative reweighted total variation (wTV) term defined as the following \cite{huang2018scale},
\begin{equation}
\begin{split}
&||\boldsymbol{f}^{(n)}||_{\text{wTV}}=\sum_{x,y,z}\boldsymbol{w}^{(n)}_{x,y,z}||\mathcal{D}\boldsymbol{f}^{(n)}_{x,y,z}||,\\
&\boldsymbol{w}^{(n)}_{x,y,z}=\frac{1}{||\mathcal{D}\boldsymbol{f}^{(n-1)}_{x,y,z}||+\epsilon},
\end{split}
\label{eq:WeightsUpdate}
\end{equation}
where $\boldsymbol{f}^{(n)}$ is the image at the $n^\text{th}$ iteration, $\boldsymbol{w}^{(n)}$ is the weight vector for the $n^\text{th}$ iteration which is computed from the previous iteration, and $\epsilon$ is a small positive value added to avoid division by zero.

 To solve the above objective function, simultaneous algebraic reconstruction technique (SART) + wTV is applied \cite{huang2018scale}. To save computation, the iterative reconstruction is initialized by $\boldsymbol{f}_{\text{U-Net}}$.

\subsection{Experimental Setup}
\begin{table}[h]
\begin{center}
\begin{scriptsize}
\begin{tabular}{|l|c|}
\hline
Parameter & Value\\
\hline
Scan angular range & $360^\circ$\\
\hline
Angular step & $1^\circ$\\
\hline
Source-to-detector distance & 1200.0\,mm\\
\hline
Source-to-isocenter distance & 600.0\,mm\\
\hline
Detector size & $600 \times 960$\\
\hline
Extended virtual detector size & $1000 \times 960$\\
\hline
Detector pixel size & 1.0\,mm $\times$ 1.0\,mm\\
\hline
Volume size & $256 \times 256 \times 256$\\
\hline
Voxel size & 1.25\,mm $\times$ 1.25\,mm $\times$ 1.0\,mm\\
\hline
\end{tabular}
\end{scriptsize}
\caption{The system configuration of cone-beam CT to validate the proposed DCR method for FOV extension.}
\label{tab:ConeBeamParametersLowDoseCTData}
\end{center}
\vspace{-10pt}
\end{table}

We validate the proposed DCR method using 18 patients' data from the AAPM Low-Dose CT Grand Challenge in cone-beam CT with Poisson noise. For each patient's data, truncated projections are simulated in a cone-beam CT system with parameters listed in Tab.~\ref{tab:ConeBeamParametersLowDoseCTData}. Poisson noise is simulated considering an initial exposure of $10^5$ photons at each detector pixel before attenuation.

For training, 425 2-D slices are chosen from 17 patients' 3-D volumes, i.\,e., picking 1 slice among every 10 slices for each patient. For test, all the 256 slices from the WCE reconstruction $\boldsymbol{f}_{\text{WCE}}$ are fed to the U-Net for evaluation. Both the training data and test data contain Poisson noise. The Hounsfield scaled images are normalized to [-1, 1] for stable training. The U-Net is trained on the above data using the Adam optimizer for 500 epochs. 
%The learning rate is set to $10^{-3}$ with a decay rate of 0.98. The $\ell_2$-norm with a parameter of $10^{-4}$ is applied to regularize the network weights. 
An $\ell_2$ loss function is used.

For reconstruction, the parameter $e_1$ is set to 0.01 for Poisson noise tolerance. A relatively large tolerance value of 0.5 is chosen empirically for $e_2$. For the wTV regularization, the parameter $\epsilon$ is set to 5\,HU for weight update. With the initialization of $\boldsymbol{f}_{\text{U-Net}}$, 10 iterations of SART + wTV only are applied to get the final reconstruction. 

\section{Results}

\begin{figure}[h]
\centering
\begin{minipage}{0.16\linewidth}
\centerline{$\boldsymbol{f}_{\text{reference}}$}
\smallskip
\end{minipage}
\begin{minipage}{0.16\linewidth}
\centerline{$\boldsymbol{f}_{\text{FBP}}$}
\smallskip
\end{minipage}
\begin{minipage}{0.16\linewidth}
\centerline{$\boldsymbol{f}_{\text{WCE}}$}
\smallskip
\end{minipage}
\begin{minipage}{0.16\linewidth}
\centerline{$\boldsymbol{f}_{\text{wTV}}$}
\smallskip
\end{minipage}
\begin{minipage}{0.16\linewidth}
\centerline{$\boldsymbol{f}_{\text{U-Net}}$}
\smallskip
\end{minipage}
\begin{minipage}{0.16\linewidth}
\centerline{$\boldsymbol{f}_{\text{DCR}}$}
\smallskip
\end{minipage}

\begin{minipage}{0.16\linewidth}
\subfigure[]
{
\includegraphics[width = \linewidth]{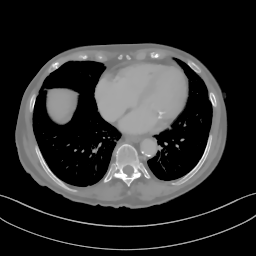}
\label{subfig:refS30}
}
\end{minipage}
\begin{minipage}{0.16\linewidth}
\subfigure[117\,HU]
{
\includegraphics[width = \linewidth]{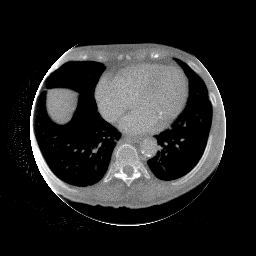}
\label{subfig:FBPS30}
}
\end{minipage}
\begin{minipage}{0.16\linewidth}
\subfigure[80\,HU]
{
\includegraphics[width = \linewidth]{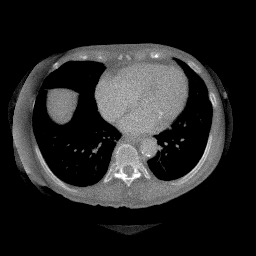}
\label{subfig:WCES30}
}
\end{minipage}
\begin{minipage}{0.16\linewidth}
\subfigure[32\,HU]
{
\includegraphics[width = \linewidth]{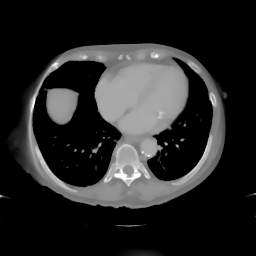}
\label{subfig:wTVS30}
}
\end{minipage}
\begin{minipage}{0.16\linewidth}
\subfigure[67\,HU]
{
\includegraphics[width = \linewidth]{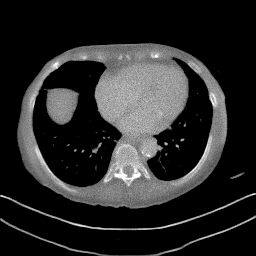}
\label{subfig:UNetS30}
}
\end{minipage}
\begin{minipage}{0.16\linewidth}
\subfigure[21\,HU]
{
\includegraphics[width = \linewidth]{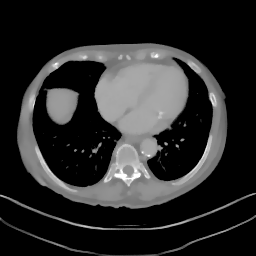}
\label{subfig:DCRS30}
}
\end{minipage}

\begin{minipage}{0.16\linewidth}
\subfigure[]
{
\includegraphics[width = \linewidth]{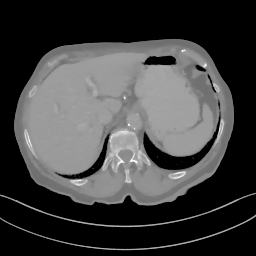}
\label{subfig:refS140}
}
\end{minipage}
\begin{minipage}{0.16\linewidth}
\subfigure[180\,HU]
{
\includegraphics[width = \linewidth]{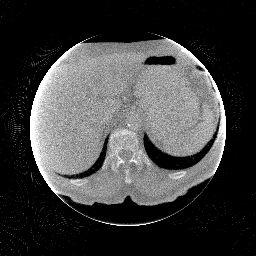}
\label{subfig:FBPS140}
}
\end{minipage}
\begin{minipage}{0.16\linewidth}
\subfigure[87\,HU]
{
\includegraphics[width = \linewidth]{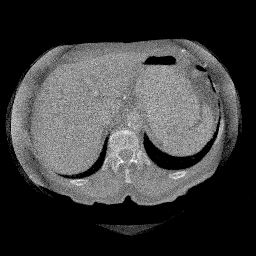}
\label{subfig:WCES140}
}
\end{minipage}
\begin{minipage}{0.16\linewidth}
\subfigure[42\,HU]
{
\includegraphics[width = \linewidth]{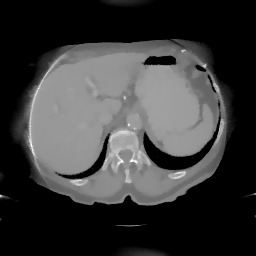}
\label{subfig:wTVS140}
}
\end{minipage}
\begin{minipage}{0.16\linewidth}
\subfigure[85\,HU]
{
\includegraphics[width = \linewidth]{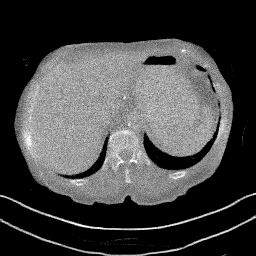}
\label{subfig:UNetS140}
}
\end{minipage}
\begin{minipage}{0.16\linewidth}
\subfigure[27\,HU]
{
\includegraphics[width = \linewidth]{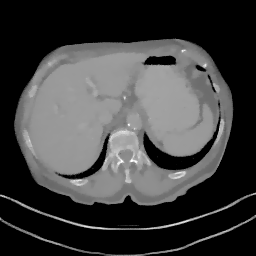}
\label{subfig:DCRS140}
}
\end{minipage}
\caption{Reconstruction results of two example slices from the test patient, window: [-600, 500]\,HU. The RMSE value inside the FOV for each method is displayed.}
\label{Fig:results}
\end{figure}

The reconstruction results of two example slices from the test patient are displayed in Fig.~\ref{Fig:results}. 
In the FBP reconstruction $\boldsymbol{f}_{\text{FBP}}$ (Figs.~\ref{Fig:results}(b) and (h)), the original FOV is observed. The anatomical structures outside this FOV are missing, while the structures inside the FOV suffer from cupping artifacts. 
WCE reconstructs certain structures outside the FOV and alleviates the cupping artifacts, according to $\boldsymbol{f}_{\text{WCE}}$ in Figs.~\ref{Fig:results}(c) and (i). However, the reconstructed structures outside the FOV is not accurate and shadow artifacts remain near the FOV boundary. 
In the wTV reconstruction $\boldsymbol{f}_{\text{wTV}}$ (Figs.~\ref{Fig:results}(d) and (j)), the cupping artifacts are mitigated. Moreover, Poisson noise is reduced as well. It achieves small root-mean-square error (RMSE) values of 32\,HU and 42\,HU for Fig.~\ref{subfig:wTVS30} and Fig.~\ref{subfig:wTVS140} inside the FOV, respectively. Nevertheless, the structures outside the FOV are still missing.
Figs.~\ref{Fig:results}(e) and (k) demonstrate that the U-Net is able to reduce the cupping artifacts and to reconstruct the anatomical structures outside the FOV as well. However, Poisson noise remains. The relative high RMSE inside the FOV indicates incorrect structures reconstructed by the U-Net.
The proposed DCR method combines the advantages of wTV and U-Net. It reconstructs the anatomical structures outside the FOV well. Meanwhile, it reduces both the cupping artifacts and the Poisson noise, as demonstrated in Figs.~\ref{Fig:results}(f) and (l). Among all the algorithms, it achieves the smallest RMSE value of 21\,HU inside the FOV.

\begin{table}[h]
\begin{center}
\begin{scriptsize}
\begin{tabular}{|l|c|c|c|c|c|}
\hline
Method & FBP & WCE & wTV & U-Net & DCR\\
\hline
%ROI1 & 118\,HU & 127\,HU & 30\,HU & 115\,HU & 25\,HU\\
%\hline
RMSE in FOV & 162\,HU & 85\,HU & 41\,HU & 77\,HU & 24\,HU\\
\hline
RMSE & 353\,HU & 179\,HU & 137\,HU & 127\,HU & 66\,HU\\
\hline
SSIM & 0.834 & 0.948 & 0.968 & 0.975 & 0.993\\
\hline
\end{tabular}
\end{scriptsize}
\caption{The quantitative evaluation results of different methods using the RMSE and SSIM metrics.}
\label{tab:quantitativeEvaluation}
\end{center}
\vspace{-10pt}
\end{table}

The average RMSE and structure similarity (SSIM) values of all the 256 slices in the test patient for different methods are displayed in Tab.~\ref{tab:quantitativeEvaluation}. DCR achieves the smallest value of 24\,HU and 66\,HU for RMSE inside the FOV and for the whole patient body, respectively. It also reaches the highest SSIM index of 0.993, which highlights the efficacy of the proposed DCR method.
 
\section{Discussion}
With deep learning prior for initialization, only a small number of iterations, e.\,g. 10 iterations in this work, are required. Therefore, it is  more efficient than conventional iterative reconstruction methods. Meanwhile, the deep learning provides information for structures outside the FOV. Therefore, it is more effective than conventional iterative reconstruction methods in the regard of FOV extension. With the integration of iterative reconstruction, it is more effective in reducing Poisson noise and more robust as well than deep learning. All in all, the proposed DCR method is a hybrid method combining the advantages of deep learning and iterative reconstruction while overcoming their shortcomings.

%\bibliographystyle{bvm2020}
%
%\bibliography{0000}

% Bitte setzen Sie hier Ihre Beitragsnummer ein und benennen Sie
% die BibTeX-Datei ebenfalls auf Ihre Beitragsnummer um.
%Kontrollzeiledef
\marginpar{\color{white}E\articlenumber} % Zeile nicht verändern!
\end{document}